\documentclass[conference]{IEEEtran}
\IEEEoverridecommandlockouts
\usepackage{cite}
\usepackage{amsmath,amssymb,amsfonts}
\usepackage{algorithmic}
\usepackage{graphicx}
\usepackage{textcomp}
\usepackage[table]{xcolor}
\usepackage{tcolorbox}
\usepackage{colortbl}
\usepackage{multirow}
\usepackage{makecell}
\usepackage{subfig}
\usepackage{tcolorbox}
\usepackage{color}
\allowdisplaybreaks[4]

\def\BibTeX{{\rm B\kern-.05em{\sc i\kern-.025em b}\kern-.08em
    T\kern-.1667em\lower.7ex\hbox{E}\kern-.125emX}}

\usepackage[pagebackref=false,breaklinks=true,colorlinks, bookmarks=false]{hyperref}
\newlength\savedwidth
\newcommand\whline{\noalign{\global\savedwidth\arrayrulewidth
		\global\arrayrulewidth 1pt}%
	\hline
	\noalign{\global\arrayrulewidth\savedwidth}}

\begin{document}

\title{AI Model Placement for 6G Networks under Epistemic Uncertainty Estimation

}

\author{Liming Huang, Yulei Wu, Juan Marcelo Parra-Ullauri, Reza Nejabati, and Dimitra Simeonidou \\
\IEEEauthorblockA{\textit{School of Electrical, Electronic and Mechanical Engineering, University of Bristol, BS8 1UB, Bristol, UK}\\
\{liming.huang, y.l.wu, jm.parraullauri, reza.nejabati, dimitra.simeonidou\}@bristol.ac.uk}
}

\maketitle

\begin{abstract}

The adoption of Artificial Intelligence (AI) based Virtual Network Functions (VNFs) has witnessed significant growth, posing a critical challenge in orchestrating AI models within next-generation 6G networks. Finding optimal AI model placement is significantly more challenging than placing traditional software-based VNFs, due to the introduction of numerous uncertain factors by AI models, such as varying computing resource consumption, dynamic storage requirements, and changing model performance. To address the AI model placement problem under uncertainties, this paper presents a novel approach employing a sequence-to-sequence (S2S) neural network which considers uncertainty estimations. The S2S model, characterized by its encoding-decoding architecture, is designed to take the service chain with a number of AI models as input and produce the corresponding placement of each AI model. To address the introduced uncertainties, our methodology incorporates the orthonormal certificate module for uncertainty estimation and utilizes fuzzy logic for uncertainty representation, thereby enhancing the capabilities of the S2S model. Experiments demonstrate that the proposed method achieves competitive results across diverse AI model profiles, network environments, and service chain requests. 
\end{abstract}


\section{Introduction}
With the deployment of 5G technology and its global standardization, many efforts are intensified to explore the future landscape of 6G networks. Virtual Network Function (VNF) placement, a key technique within Network Function Virtualization (NFV) for 5G, is poised to retain its crucial role in shaping the architecture of 6G networks. Different from traditional VNFs, typically reliant on specific functional software, the evolving trend in 6G networks is the increasing integration of VNFs driven by artificial intelligence (AI) models~\cite{10090468}. AI model-based VNFs possess a significant advantage over traditional software-based VNFs due to their adaptability and efficiency\footnote{In this paper, we refer to AI model-based VNFs as AI models, and to traditional software-based VNFs as traditional VNFs.}. Unlike traditional VNFs, AI models can dynamically adjust to varying network conditions in real-time, optimizing performance and resource allocation. Additionally, AI models have the capability to self-learn and improve over time, enhancing overall network functionality and reducing the need for manual intervention. This superior adaptability and self-optimization make AI models a more reliable and scalable solution for modern network infrastructures, ultimately leading to improved reliability, performance, and cost-effectiveness.

A significant amount of research has contributed to traditional VNF placement, as evidenced by studies such as~\cite{8945291,8466627,9311156}. Due to their typically fixed power consumption, running time, and model size, traditional VNFs are usually placed under conditions with unchanged and predictable performance factors. 
In contrast, AI model placement presents serious challenges due to the inherent uncertainties surrounding AI models. One of the primary challenges is the variability in computing resource requirements of AI models. AI models typically need to be trained, and the duration of training can be difficult to predict due to a variety of factors such as the amount and quality of data used, parameter settings, and the computational resources available. These unpredictable training times can lead to uncertainty in terms of computing resource allocation and usage when placing AI models.
Moreover, the storage space required for training AI models varies significantly based on the model's architecture and the input data structure. This uncertainty impacts the allocation and management of storage resources within the network. 
In addition, the operational performance of AI models fluctuates with changing network conditions and evolving data patterns, hindering predictions of their real performance.
Consequently, these uncertainties surrounding computing resources, model storage, and model performance pose significant hurdles in devising optimal placement strategies for AI models within dynamic network environments. 
In this paper, we refer to these uncertainties as epistemic uncertainties (derived from the Greek word ``episteme'', which signifies ``knowledge'')~\cite{tagasovska2019single}, as existing placement models lack the necessary understanding to address these undetermined variables adequately.

To this end, this paper focuses on AI model placement, considering various epistemic uncertainties inherent in AI models. To identify the optimal placement strategy that satisfies constraints from network environments and Service Level Agreements (SLA), we formulate this task as a constrained combinatorial optimization problem.
To achieve AI model placement, we devise a sequence-to-sequence (S2S) neural network with an encoder-decoder structure. For a Network Service (NS) request, the proposed S2S neural network can consider the state of the NFV infrastructure and then learn a placement policy to minimize the overall energy consumption while satisfying specific network constraints.
Since it is challenging to directly obtain optimal solution labels, this paper adopts a neural combinatorial optimization (NCO) framework~\cite{bello2016neural} to address the AI model placement task. The NCO framework is devised by training the neural network within a reinforcement learning (RL) environment, where the RL reward can help the neural network perform loss design and backpropagation without the need for supervised labels. 
To solve the uncertainty issues, we propose an uncertainty estimator where the Orthonormal Certificate (OC) module~\cite{tagasovska2019single} is used for uncertainty estimations, and a Gaussian fuzzy logic is employed to fuse the uncertainty estimations into the neural network model.

The main contributions of this paper are given as follows:
\begin{itemize}
	\item This paper is the first of its kind to address AI model placement problems for 6G networks, explicitly considering the uncertainties introduced by AI models due to varying computing resource consumption, dynamic storage requirements, and changing model performance.
	\item A S2S neural network with uncertainty estimation is devised for this problem. The OC module is proposed for uncertainty estimations and a Gaussian fuzzy logic is devised to integrate the uncertainty estimations into the S2S model.
    \item Experiments demonstrate the competitive performance of our model across a spectrum of AI model profiles, varying network environments, and diverse service chain requests. Notably, in scenarios where a service chain comprising 30 AI models is allocated to 20 servers, our model excels with a remarkable request acceptance ratio that is 3 $\sim$ 4 times more than that of other related models.
\end{itemize}

\section{Related Work}

\subsection{VNF Placement}

The VNF placement problem has garnered considerable attention. Woldeyohannes et al.~\cite{8466627} addressed it through a multi-objective integer linear programming approach, optimizing admitted flows, node and link utilization, and meeting delay requirements. Chen et al.~\cite{9311156} focused on delay-aware VNF placement, introducing heuristic algorithms for optimal backup VNF placement and routing to ensure NS availability while minimizing transmission delays. Studies on VNF reusability and sharing for cost reduction have been conducted as well. Malandrino et al.~\cite{8874992} emphasized VNF placement within a single physical node with shared instances among multiple NSs, using an efficient heuristic to prioritize NSs while adhering to delay requirements. Ren et al.~\cite{9050847} examined VNF reusability in delay-aware multicast NS placement, developing algorithms to maximize system throughput and minimize costs for admitted multicast NSs while meeting delay requirements.

While the aforementioned research has made significant strides in addressing VNF placement problems, it predominantly focuses on traditional VNFs and overlooks the increasingly prevalent usage of AI models. Moreover, the uncertainties associated with AI model placement remain unexplored within this body of work. This paper aims to address this important gap by investigating the implications of deploying AI models, a dimension largely overlooked in prior studies on VNF placement.

\subsection{Uncertainty Learning}

Uncertainty learning holds a key role in determining when to abstain from predictions across various tasks. In deep learning, Natasa et al.~\cite{tagasovska2019single} introduced neural network estimations for aleatoric and epistemic uncertainties. In RL, Vincent et al.~\cite{mai2022sample} proposed the utilization of uncertainty estimation to alleviate the adverse effects of noisy supervision, and introduced an inverse variance RL approach to enhance sample efficiency and overall performance.

In the field of VNF placement, existing research primarily focuses on dynamic resource allocation in uncertain network environments~\cite{8030590, 9348044}. However, there has been limited attention given to understanding uncertainty issues arising from the training process of AI model-based VNFs. This paper uniquely centers on AI-based VNF placement and explores an uncertainty estimator that addresses epistemic uncertainty through the OC module and fuzzy logic. Overall, this paper contributes to a more comprehensive understanding of optimal AI model placement strategies in NFV.

\section{Problem Formulation}
\label{sec:PF}

\subsection{Problem Statement}

This paper addresses the strategic placement of a variety of AI models into a set of host servers, in support of the delivery of network services. Each host server possesses limited computing, storage, and connectivity resources. The host servers are interconnected using specific link connections, each with its unique attributes such as bandwidth and propagation delay. 

The primary objective of our problem formulation is to optimize the AI model placement involved in a given network service, minimizing the overall infrastructure power consumption. This objective holds significant importance, as it directly impacts operational costs, promotes sustainability by reducing the carbon footprint, extends hardware lifespan, and ensures long-term scalability. Beyond minimizing power consumption, this strategic placement must align with constraints concerning physical resources availability, link capacities, and the latency thresholds mandated by individual service requests. More importantly, the uncertainty factors brought by AI models should be considered in solving this problem.

In this paper, the set of host servers is denoted as $\{ h_1, h_2, \cdots, h_n \} \in H$, and the repository of available AI models is represented as $\{ a_1, a_2, \cdots \} \in A$. A network service comprises $m$ AI models forming a service chain $s = (a_1, a_2, \cdots, a_m) \in S$. This problem entails identifying the optimal set of placements, denoted as $x \in \{0, 1\}^{m \times n}$, where $x_{ah}$ represents a binary status variable indicating whether an AI model $a \in s$ is placed in host $h \in H$ (1 for positive placement and 0 for negative). The search space is denoted as $\Omega = { x \in \{ 0, 1 \}^{m \times n}\ s.t.\ \sum\nolimits_h {{x_{ah}} = 1},\ \forall a \in s }$, indicating that each AI model can only be placed in one host server at a time. 
In order to more accurately restore the real network environment, we introduce server activation variables $y_h \in \{0, 1\}$ indicating whether the server is executing any AI models (1 for active, and 0 for powered off) and link activation variables $g_i \in \{0, 1\}$ signifying whether a link carries traffic ($1$) or not ($0$). 

For the computing power consumption, each unit of CPU and GPU in activated servers ($y_h = 1$) consumes $W_h^{c}$ and $W_h^{g}$ watts, respectively, and their power consumption increases linearly with the cumulative CPU and GPU demand of the placed AI models~\cite{8945291}. Additionally, we consider the standby power of activated servers to be $W_h^{e}$.
Regarding power consumption in the links, it is determined by multiplying the bandwidth utilized in each link by the consumption of each bandwidth unit $W_{net}$.
The computing resource availability of each server $h$ is denoted as $r_h^{type} \in R$, and the resource requirement of an AI model $a$ is denoted as $r_a^{type}$, where $type \in \{c,g\}$ with $c$ representing CPU and $g$ representing GPU. The bandwidth required for data transfer of AI model $a$ within the service $s$ is denoted as $b_a^s$. The maximum allowed bandwidth for each link $i$ is represented as $b_i$. Regarding latency requirements, $l_a$ signifies the latency due to the computation time of AI model $a$, while $l_i^s$ represents the latency in link $i$ due to service $s$. The maximum latency permitted for each service chain $s$ is denoted as $l_s$. A summary of defined variables and parameters is presented in Table~\ref{table_para}.

\begin{table}
	\centering
	\caption{
		Problem Formalization Variables
	}
	\label{table_para}
	\setlength\tabcolsep{1.0em}
	\renewcommand{\arraystretch}{1.2}
	\rowcolors{2}{white}{gray!20}
    \resizebox{1.0\columnwidth}{!}{
	\begin{tabular}{l|ll}
		\whline
		1     & $H / L / A / S / R$     & set of hosts / links / AI models / NS chains / resources \\
		2     & $r_{h}$    & amount of resources $r$ available in host $h$ \\
		3     & $r_{a}$     & amount of resources $r$ requested by AI model $a$ \\
		4     & $W_{h}^{e}$     & idle power consumption of host $h$ \\
		5     & $W_{h}^{c}$ / $W_{h}^{g}$     & power consumption of each CPU/GPU in host $h$ \\
		6     & $W_{net}$     & power consumption per bandwidth unit on links \\
		7     & $b_{i}$     & bandwidth of the link $i$ \\
		8     & $l_{a}$     & latency due to computation time of AI model $a$ \\
		9     & $l_{i}^s$     & latency on the link $i$ produced by service chain $s$ \\
		10     & $b_{a}^s$     & bandwidth demanded by $a$ in service chain $s$ \\
		11     & $l^s$     & maximum latency allowed on the service chain $s$\\
        12     & $d_h$     & amount of disk spaces $d$ available in host $h$\\
        13     & $d_a$     & amount of disk spaces $d$ requested by AI model $a$\\
        14     & $q_{sla}$      & model performance required by SLA\\
        15     & $q_a$     & model performance achieved by AI model $a$\\
		16     & $x_{ah}$     & binary placement variable for AI $a$ in host $h$  \\
		17     & $y_{h}$     & binary activation variable for host $h$ \\
		18     & $g_{i}$     & binary activation variable for link $i$ \\
		\whline
	\end{tabular} }
\end{table}

\subsection{Mathematical Formulation} \label{sec:MF}

In previous traditional VNF placement~\cite{agarwal2018joint, pei2019optimal}, host servers are usually characterized by a certain power profile that grows in proportion to the computing utilization because traditional VNFs usually have fixed energy consumption. Therefore, the optimized cost function with certain relationships for traditional VNF placement can be described as follows:
\begin{align}
	&\mathop {\arg \min }\limits_{x \in \Omega } \Big \{ \sum\limits_{h \in H} { \Big [  \sum\limits_{a \in s} { ({W_h^{c}} \cdot {r_a^{c}} + {W_h^{g}} \cdot {r_a^{g}}) \cdot {x_{ah}} }  + W_h^{e} \cdot {y_h}  \Big ] } \nonumber \\ 
	&+ \sum\limits_{i \in L} { \sum\limits_{a \in s} {W_{net}} \cdot {b_a^s} }  \cdot {x_{ah}} \Big \} \label{eq_VNF_1} \\ 
	&s.t. \nonumber \\
	&\sum\limits_{a \in s} {{r_a} \cdot {x_{ah}}}  \le {y_h} \cdot {r_h},~\forall r_a \in r_a^{type}, r_h \in r_h^{type}, h \in H \label{eq_VNF_2}\\ 
	&\sum\limits_{a \in s} {b_a^s \cdot {x_{ah}}}  \le {g_i} \cdot {b_i},~\forall i \in L \label{eq_VNF_3}\\ 
	&\sum\limits_{h \in H} {\sum\limits_{a \in s} {{l_a} \cdot {x_{ah}}} }  + \sum\limits_{i \in L} {\sum\limits_{a \in s} {l_i^s} }  \cdot {x_{ah}} \le {l^s},~\forall h \in H, i \in L \label{eq_VNF_4} 
\end{align}

In the above formulation, Eq.~(\ref{eq_VNF_1}) represents the computing power consumption attributed to activated servers and the collective consumption associated with active links. Eqs.~(\ref{eq_VNF_2})-(\ref{eq_VNF_4}) represent the constraints imposed by computing resources, bandwidth resources, and latency requirements.

AI models, unlike traditional VNFs, typically require a training process to achieve specific functions, leading to various uncertainties. Firstly, the model training time $t$ is uncertain during AI model training, resulting in dynamic computing resource consumption. Variations in input data, network environment, and computing resources cause considerable variability in the model's training time. Even in a static environment with fixed parameter settings and using the same data to train the same model, the training time of the AI model may vary due to the stochastic nature of gradient descent in neural networks.
Secondly, the utilization of storage space $d$ is uncertain in the model training process. For example, an AI model typically converges after 100 training epochs, and we usually save the trained model at that epoch for later use. However, in another training process for the same model, convergence does not occur at the 100-th epoch. As a result, the model will be stored from the 100-th epoch until the actual convergence epoch, as we have empirically set the model to save from the 100-th epoch onwards. Consequently, the storage space utilization increases for this training process. 
Thirdly, the model performance quality $q$ is uncertain. Due to the stochastic nature of gradient descent in neural networks, the optimal point achieved by the model in each training instance is unlikely to be identical, even though they may be close, resulting in uncertain model performance. Therefore, in order to conduct proper AI model placement, these three uncertainties must be taken into account: computing resource consumption, storage space usage $d$, and model performance quality $q$. 

Incorporating the aforementioned uncertainties into the placement strategy of traditional VNFs, as shown in Eqs.~(\ref{eq_VNF_1})-(\ref{eq_VNF_4}), we can formulate the optimization of AI model placement as follows:
\begin{align}
	&\mathop {\arg \min }\limits_{x \in \Omega } \Big \{ \sum\limits_{h \in H}  \Big [  \sum\limits_{a \in s} { \big({W_h^{c}} \cdot {r_a^{c}} \cdot f_c + {W_h^{g}} \cdot {r_a^{g}} \cdot f_g\big) \cdot {x_{ah}}  }    \nonumber \\ 
	&+ W_h^{e} \cdot {y_h}  \Big ] + \sum\limits_{i \in L} { \sum\limits_{a \in s} {W_{net}} \cdot {b_a^s} }  \cdot {x_{ah}} \Big \} \label{eq_AI_1} \\ 
	&s.t. \nonumber \\
	&\sum\limits_{a \in s} {{r_a^c} \cdot {x_{ah}} \cdot f_c }  \le {y_h} \cdot {r_h^c},~\forall  h \in H \label{eq_AI_2}\\ 
        &\sum\limits_{a \in s} {{r_a^g} \cdot {x_{ah}} \cdot f_g }  \le {y_h} \cdot {r_h^g},~\forall  h \in H \label{eq_AI_2.2}\\ 
	&\sum\limits_{a \in s} {b_a^s \cdot {x_{ah}}}  \le {g_i} \cdot {b_i},~\forall i \in L \label{eq_AI_3}\\ 
	&\sum\limits_{h \in H} {\sum\limits_{a \in s} {{l_a} \cdot {x_{ah}}} }  + \sum\limits_{i \in L} {\sum\limits_{a \in s} {l_i^s} }  \cdot {x_{ah}} \le {l^s},~\forall h \in H, i \in L \label{eq_AI_4} \\
	&\sum\limits_{a \in s} {{d_a} \cdot {x_{ah}} \cdot f_d}  \le {y_h} \cdot {d_h},~\forall h \in H \label{eq_AI_5} \\
	&\sum\limits_{a \in s} {{q_a} \cdot {x_{ah}} \cdot f_q}  \geq q_{sla},~\forall h \in H \label{eq_AI_6}
\end{align}
where $d_h$ denotes the available storage space in host servers, and $q_{sla}$ is the model performance specified by the SLA. $f_c$ and $f_g$ are uncertainty functions for CPU and GPU power consumption, respectively. $f_d$ is the uncertainty function for storage space usage. $f_q$ represents the model performance uncertainty function.

\section{Optimizing AI Model Placement with Uncertainty Estimation}

\begin{figure}
	\centering
	\includegraphics[width=0.90\linewidth]{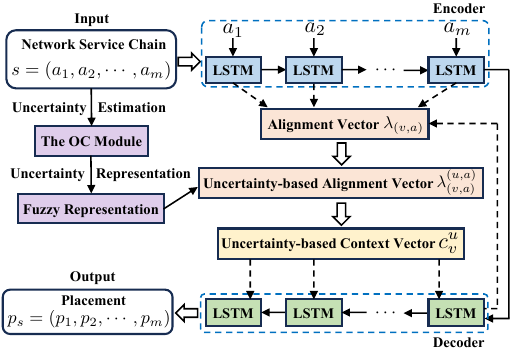}
	\caption{The framework of our AI placement model.}
	\label{fig:figure-model}
\end{figure}

\subsection{Sequence-to-Sequence Model for Placement}
\label{sec:s2s}

To facilitate AI model placement, we employ a sequence-to-sequence (S2S) model following~\cite{sutskever2014sequence, 8945291}, which is capable of optimizing sequences with dynamic lengths. Given an NS chain $s = (a_1, a_2, \cdots, a_m)$ comprising a sequence of AI models as the input, our goal is to determine the optimal server positions $p_s = (p_1, p_2, \cdots, p_m)$ for each AI model, while achieving the main objective outlined in Eq.~(\ref{eq_AI_1}) and adhering to constraints specified in Eqs.~(\ref{eq_AI_2})-(\ref{eq_AI_6}).

The S2S model utilizes an encoder-decoder architecture as illustrated in Fig.~\ref{fig:figure-model}. For the encoder part, the stacked Long Short-Term Memory (LSTM) cells are used to process the input sequence, generating the source hidden state ${\overline \rho_a}$ for each AI model $a$. For the decoder part, we use an attention-based LSTM model following~\cite{bahdanau2014neural}. The decoder model has the same step numbers as the input sequence size, and outputs the host position to place the AI model introduced by the encoder part for each step. 
The decoder's hidden state for the step $v$ is denoted as $\rho_v = f(\rho_{v-1}, \overline \rho_{v-1}, c_v)$, which is a function by combining the previous state with a context vector $c_v$. Specifically, the context vector $c_v$ is calculated as follows:
\begin{equation} \label{eq:context_v}
	{c_v} = \sum\nolimits_{a \in A} {{\lambda_{(v,a)}}} \cdot {\overline \rho_a},
\end{equation}
where $\lambda_{(v,a)}$ represents the alignment score for each source hidden state ${\overline \rho_a}$ in the decoding process. The alignment score $\lambda_{(v,a)}$ is calculated using the current target hidden state $\rho_v$ with each source hidden state ${\overline \rho_a}$ as follows:
\begin{equation} \label{eq:align_score}
	\lambda _{(v,a)} = \text{softmax} \big (\omega \tanh ({\xi_v}{\rho_v} + {\xi_a}{\overline \rho_a}) \big ),
\end{equation}
where $\omega$, ${\xi_v}$, and ${\xi_a}$ are weight matrices that will be learned from the neural network. 

\subsection{Uncertainty Estimation and Combination} 

\subsubsection{\textbf{Uncertainty Estimation}}

As outlined in $\S$\ref{sec:MF}, AI models consist of a variety of uncertain factors that necessitate consideration in their placement. In this paper, we introduce the Orthonormal Certificate (OC) module~\cite{tagasovska2019single} into the S2S model. The OC module is an epistemic uncertainty estimator, which works on deep neural networks by mapping the input features to zero and minimizing the OC-based loss function for uncertainty estimation. For a comprehensive understanding of OC, detailed technical information can be referenced in~\cite{tagasovska2019single}.

This study investigates four uncertainty factors affecting AI model placement: CPU consumption, GPU consumption, disk utilization, and model performance. In the S2S encoder part, the embedding feature for the input chain is denoted as ${\overline \rho} \in {{\mathbb{R}}^{m \times e }} $, where $m$ represents the chain's length and $e$ is the embedding size of the S2S model. To incorporate the OC module for uncertainty estimation, we first concatenate the embedding feature ${\overline \rho}$ with information regarding CPU consumption, GPU consumption, disk utilization, and model performance for AI models to generate the OC input feature ${\overline x_\rho} \in {{\mathbb{R}}^{m \times (e + 4)}}$. Then, we input the feature ${\overline x_\rho}$ into the OC module to generate uncertainty estimation as follows:
\begin{equation} \label{eq_ue}
	{u_e}({\overline x_\rho})  = \text{OC}({\overline x_\rho}, d_u),
\end{equation}
where $\text{OC}(*)$ represents the linear orthonormal certificate module proposed in~\cite{tagasovska2019single}, $d_u$ is the number of the uncertainty factors where we set $d_u = 4$ in this study, and ${u_e}(*)$ is the uncertainty estimation function.

\subsubsection{\textbf{Fuzzy Representation for Uncertainties}}

To integrate the uncertainty estimations into the S2S model, we propose the use of fuzzy logic to create a fuzzy layer to represent uncertainties. Specifically, for a model $a$ in an NS chain $s = (a_1, a_2, \cdots, a_m)$ and its OC feature ${\overline x_a} \in {\overline x_\rho}$, the uncertainty estimation is denoted as $u_e({\overline x_a})$, and the fuzzy uncertainty representation is denoted as $F \big\{u_e({\overline x_a}) \big\}$. In this paper, we employ Gaussian fuzzy logic~\cite{wu2012twelve} to represent the uncertainties. Given the uncertainty estimation $u_e({\overline x_a})$, its Gaussian fuzzy representation can be calculated as follows:
\begin{equation} \label{eq:fuzzy-layer}
	F \big\{u_e({\overline x_a}) \big\} = Exp\Big\{- \frac{{\big(u_e({\overline x_a}) - \mu \big)}^{2}}{\sigma^{2}}\Big\},	
\end{equation}
where $\mu$ is the mean value of the Gaussian membership function, and $\sigma^{2}$ is the variance set according to~\cite{7482843}.

\subsubsection{\textbf{Uncertainty Combination}}

In $\S$\ref{sec:s2s}, we construct an attention-based context vector $c_v$, which serves as the enhanced feature for each input $a$. To incorporate the uncertainties of AI models into the placement model, we propose to integrate the fuzzy representation of uncertainties, as shown in Eq.~(\ref{eq:fuzzy-layer}), with the attention alignment score $\lambda _{(v,a)}$, as shown in Eq.~(\ref{eq:align_score}), to generate the uncertainty-based alignment score $\lambda_{(v,a)}^{(u,a)}$. Specifically, the combination strategy is given as follows:
\begin{equation} \label{eq:fuzzy-combine}
	\lambda_{(v,a)}^{(u,a)} = \text{softmax} \Big (\omega \tanh ({\xi_v}{\rho_v} + {\xi_a}{\overline \rho_a}) \cdot F \big\{u_e({\overline x_a}) \big\} \Big ).
\end{equation}

Then, we update the context vector $c_v$ to the uncertainty-based context vector $c_v^u$ as follows:
\begin{equation} \label{eq:context_u}
	{c_v^u} = \sum\nolimits_{a \in A} {{\lambda_{(v,a)}^{(u,a)}}} \cdot {\overline \rho_a}.
\end{equation}
This approach allows the uncertainty characteristics of AI models to be incorporated into the S2S model. The combined vector ${c_v^u}$ is then input into the decoder part to generate the placement position for each AI model.

\begin{figure}
	\centering
	\includegraphics[width=1.0\linewidth]{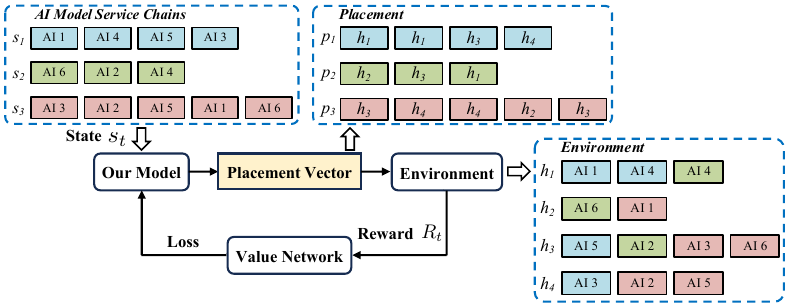}
	\caption{The learning framework for the S2S model with RL environment.}
	\label{fig:figure-rl}
\end{figure}

\subsection{Neural Combinatorial Optimization}
Since obtaining supervised labels for loss design is challenging in this task, we adopt an RL framework for model training as shown in Fig.~\ref{fig:figure-rl}. Given an NS chain as the state, the S2S model is used to generate the placement vector. Subsequently, the RL environment, incorporating network constraints as depicted in Eqs.~(\ref{eq_AI_2})-(\ref{eq_AI_6}), provides rewards for placement vectors. Furthermore, a value network, comprising the S2S model and a multi-layer perception (MLP) layer, is utilized to receive RL rewards and generate the baseline for the loss design of the S2S model, as described in~\cite{8945291}.

\begin{table}[!t]
	\centering
	\caption{
		\small Network Infrastructure  Environment Settings
	}
	\label{table_host}
	\setlength\tabcolsep{0.3em}
	\renewcommand{\arraystretch}{1.1}
	\resizebox{0.9\columnwidth}{!}{
	\begin{tabular}{l|cccccccccc}
		\multicolumn{11}{l}{Network Environment with 10 Host Servers:} \\
		\hline
		Host No. & 1     & 2     & 3     & 4     & 5     & 6     & 7     & 8     & 9     & 10 \\
		\hline
		CPU / GPU Capacity & 10    & 9     & 8     & 7     & 6     & 6     & 6     & 6     & 6     & 6 \\
		Disk Size (GB) & 10    & 10     & 10     & 8     & 8     & 8     & 6     & 6     & 6     & 6 \\
		Link Bandwidth (Mbps) & 1000  & 1000  & 500   & 400   & 300   & 300   & 300   & 300   & 300   & 300 \\
		Link Latency (ms) & 30    & 50    & 10    & 50    & 50    & 50    & 50    & 50    & 50    & 50 \\
		\hline
		\multicolumn{11}{l}{\multirow{2}[2]{*}{Network Environment with 20 Host Servers:}} \\
		\multicolumn{11}{c}{} \\
		\hline
		Host No. & 1     & 2     & 3     & 4     & 5     & 6     & 7     & 8     & 9     & 10 \\
		\hline
		CPU / GPU Capacity  & 10    & 10     & 9     & 9     & 8     & 8     & 7     & 7     & 6     & 6 \\
		Disk Size (GB)        & 10    & 10     & 10     & 10     & 10     & 10     & 8     & 8     & 8     & 8 \\
		Link Bandwidth (Mbps) & 1000  & 1000  & 1000   & 1000   & 500   & 500   & 400   & 400   & 300   & 300 \\
		Link Latency (ms)    & 30    & 30    & 50    & 50    & 10    & 10    & 50    & 50    & 50    & 50 \\
		\hline
		Host No.              & 11     & 12     & 13     & 14     & 15     & 16     & 17     & 18     & 19     & 20 \\
		\hline
		CPU Capacity          & 6    & 6     & 6     & 6     & 6     & 6     & 6     & 6     & 6     & 6 \\
		Disk Size (GB)        & 8    & 8     & 6     & 6     & 6     & 6     & 6     & 6     & 6     & 6 \\
		Link Bandwidth (Mbps) & 300  & 300  & 300   & 300   & 300   & 300   & 300   & 300   & 300   & 300 \\
		Link Latency (ms)    & 50    & 50    & 50    & 50    & 50    & 50    & 50    & 50    & 50    & 50 \\
		\hline
	\end{tabular}	}
\end{table}

\begin{table}[!t]
	\centering
	\caption{
		\small AI Model Profiles
	}
	\label{table_AI}
	\setlength\tabcolsep{0.20em}
	\renewcommand{\arraystretch}{1.2}
	\resizebox{1\columnwidth}{!}{
	\begin{tabular}{l|cccccccc}
		\hline
		AI Model No. & 1     & 2     & 3     & 4     & 5     & 6     & 7     & 8    \\
		\hline
		Required CPU / GPU Capacity & 4+$\Delta_1$    & 3+$\Delta_2$     & 3+$\Delta_3$     & 2+$\Delta_4$      & 2+$\Delta_5$      & 2+$\Delta_6$     & 1+$\Delta_7$      & 1+$\Delta_8$      \\
		Required Disk Size (GB) & 4+$\Theta_1$   & 4+$\Theta_2$    & 3+$\Theta_3$    & 3 + $\Theta_41$    & 2+$\Theta_5$    & 2+$\Theta_6$    & 1+$\Theta_7$    & 1+$\Theta_8$   \\
		Required Bandwidth (Mbps) & 100  & 80  & 60   & 20   & 20   & 20   & 20   & 20   \\
		Required Latency (ms) & 100    & 80    & 60    & 20    & 20    & 20    & 20    & 20   \\
		Task Completion Rate (\%) & 80+$\Upsilon_1$    & 80+$\Upsilon_2$    & 80+$\Upsilon_3$    & 80+$\Upsilon_4$    & 80+$\Upsilon_5$    & 80+$\Upsilon_6$    & 80+$\Upsilon_7$    & 80+$\Upsilon_8$   \\
		\hline
	\end{tabular}	}
\end{table}

\section{Evaluation}
\label{sec:evaluation}

\subsection{Experimental Settings}

To evaluate the proposed model, this paper selects the network infrastructures with two different sizes consisting of 10 host servers and 20 host servers, respectively. 
The detailed settings of the network infrastructure environment are provided in Table~\ref{table_host}. The AI model profiles are given in Table~\ref{table_AI}, where $\Delta_i$, $i \in \{ 1, 2, \cdots, 8 \}$, $\Theta_j$, $j \in \{ 1, 2, \cdots, 8 \}$ and $\Upsilon_k$, $k \in \{ 1, 2, \cdots, 8 \}$ are uncertain variables randomly generated from two distributions, namely the normal distribution and the uniform distribution. Power consumption parameters in Eq.~(\ref{eq_AI_1}) are specified as follows: $W_h^c = W_h^g = 200$, $W_h^e = 100$ and $W_{net} = 0.1$. As for model parameters, the learning rate is set to $0.0001$, the embedding size is set to $10$ according to~\cite{sutskever2014sequence}. 
The source code is available on the project's homepage\footnote{https://github.com/lmhuang-me/AI-Placement.git}.

\subsection{Comparison Baselines and the Evaluation Metrics}

\textbf{Baseline Models:} We evaluate our method with three baseline models: NCO~\cite{8945291}, Gecode solver~\cite{schulte2006gecode} and the First Fit (FF) heuristic algorithm~\cite{kumaraswamy2019bin}. NCO is a deep RL method for VNF placement without considering uncertainties. The Gecode solver is an algorithm based on the classical Branch and Bound paradigm. The FF algorithm is designed with a heuristic first fit strategy.

\textbf{Evaluation Metric:} The evaluation metric used in the experiment is the NS request acceptance ratio, defined as the ratio of successfully placed requests to the total number of requests. For example, if 100 requests are received and 50 are successfully placed, then the acceptance ratio is $50\%$.

\subsection{Model Performance with Different Requests}

To evaluate the proposed model, we generate 64 NS requests and 128 NS requests for placement. In the network environment with 10 hosts, we evaluate NS chains consisting of 12, 14, 16, and 18 AI models sequentially. In the network environment with 20 hosts, we select NS chains consisting of 20, 24, 28, and 30 AI models. From the results in Table~\ref{table_request}, it is evident that our model achieves more than $90\%$ successful placement for AI models with the considered uncertainties at $\langle$ Length 12, Host number 10 $\rangle$ and $\langle$ Length 20 and 24, Host number 20 $\rangle$ for both 64 and 128 requests. This illustrates that our model has strong placement capabilities for AI models with uncertain properties. In the next section, we conduct comparative experiments between the proposed method and baseline models to further evaluate the performance of our model.

\begin{table}[!t]
	\centering
	\caption{
		\small Experimental Results of Our Method with Different Number of NS Requests, Hosts, and AI Models of a Service Request}
	\label{table_request}
	\setlength\tabcolsep{0.20em}
	\renewcommand{\arraystretch}{1.2}
	\resizebox{1\columnwidth}{!}{
		\begin{tabular}{c|c|cc|c|c|cc}
			\whline
			\makecell[c]{No. of \\NS Requests} & \makecell[c]{No. of \\Hosts} & \makecell[c]{Service\\ Length } & \makecell[c]{NS Request\\ Accept Ratio} & \makecell[c]{No. of \\NS Requests} & \makecell[c]{No. of \\Hosts} & \makecell[c]{Service\\ Length} & \makecell[c]{NS Request\\ Accept Ratio}  \\
			\hline
			\multirow{8}{*}{64} & \multirow{4}{*}{10} & 12    & 98.4\%     & \multirow{8}{*}{128} & \multirow{4}{*}{10} & 12    & 97.6\% \\
			&       & 14    & 85.9\%     &       &       & 14    & 82.0\% \\
			&       & 16    & 46.9\%     &       &       & 16    & 43.8\% \\
			&       & 18    & 12.5\%     &       &       & 18    & 12.5\% \\
			\cline{2-4}\cline{6-8}      & \multirow{4}{*}{20} & 20    & 95.3\%     &       & \multirow{4}{*}{20} & 20    & 95.3\% \\
			&       & 24    & 93.7\%     &       &       & 24    & 92.2\% \\
			&       & 28    & 70.3\%     &       &       & 28    & 68.7\% \\
			&       & 30    & 35.9\%     &       &       & 30    & 37.5\% \\
			\whline
		\end{tabular}%
		}
\end{table}

\begin{figure}[!t]
	\centering
	\subfloat[]{
		\includegraphics[width=0.46\linewidth]{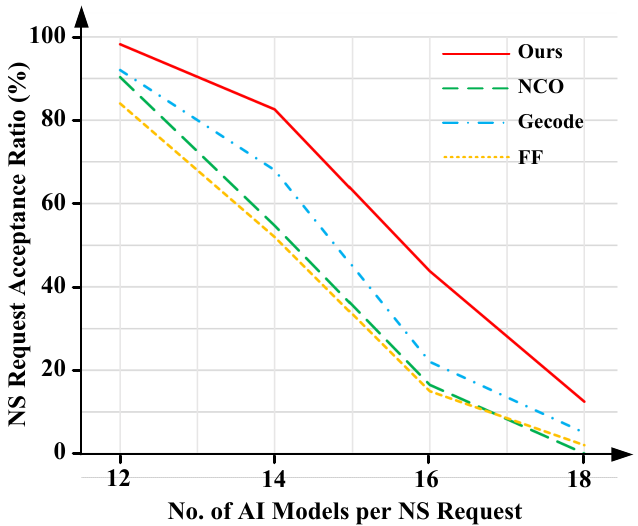}
		\label{fig:comparison1}
	}
	\hfil
    \vspace{-5pt}
	\subfloat[]{
		\includegraphics[width=0.46\linewidth]{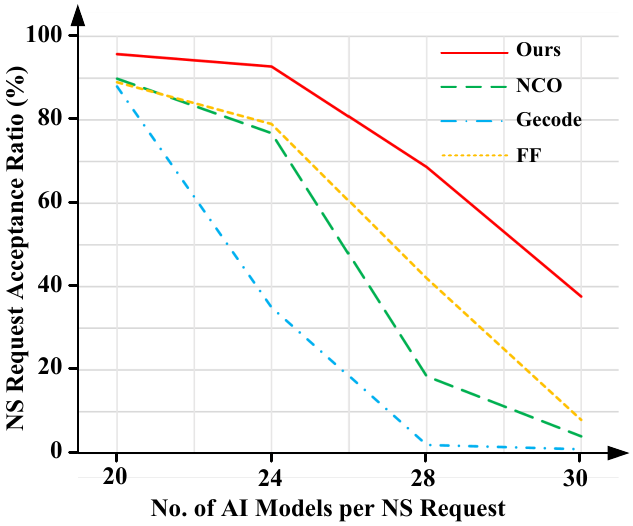}
		\label{fig:comparison2}
	}
	\caption{
		Comparisons with NCO~\cite{8945291}, Gecode~\cite{schulte2006gecode} and FF~\cite{kumaraswamy2019bin} with 128 NS requests in (a) 10 hosts and (b) 20 hosts.
	}
	\label{fig:comparison}
\end{figure}

\subsection{Comparison with Baseline Models}

To further validate the proposed approach, we compare our model with NCO~\cite{8945291}, Gecode solver~\cite{schulte2006gecode} and the FF algorithm~\cite{kumaraswamy2019bin}. The experiment is conducted using 128 NS requests, and the results are shown in Fig.~\ref{fig:comparison}. The results demonstrate that the NS request accept ratio of our method is significantly superior to that of the other three methods when handling NS requests with varying numbers of AI models. This superiority arises because our model incorporates estimates of AI model uncertainties into resource allocation.

\section{Conclusion}
\label{sec:conclusion}

This paper addressed the critical challenge of AI model placement in next-generation networks, particularly in the context of 6G, where VNFs heavily rely on AI implementation. The inherent uncertainties introduced by AI models, including varying computing resource requirements, unpredictable storage needs and changing performance, make optimal AI model placement more complex compared to traditional software-based VNFs. 
To tackle this problem, we proposed an S2S neural network with uncertainty estimation for AI model placement. To handle the uncertainties introduced by the AI model training process, we incorporated the OC module for predicting uncertainties and utilized fuzzy logic for uncertainty representations. These uncertainty estimations are integrated into the context vector of the S2S model, thereby optimizing the placement strategy for AI models. 
Our experimental results demonstrate the effectiveness of the proposed strategy, showcasing competitive performance in addressing the challenges associated with AI model placement under uncertainties. This research contributes valuable insights to the ongoing development of next-generation networks by providing a robust and adaptive solution for optimal AI model placement.

\section*{Acknowledgements}
This work has been partially sponsored by  the UK GOV DSIT (FONRC) project REASON.

{
	\bibliographystyle{IEEEtran}
	\bibliography{ICC}
}

\end{document}